%% file: covid_politicalregimes.tex
\documentclass[a4paper,12pt]{article}
\usepackage[T1]{fontenc}
\usepackage[english]{babel}
\usepackage{geometry}
\usepackage{graphicx}
\usepackage{tabularx}
\newcolumntype{x}{>{\centering\arraybackslash}X} 
\usepackage{booktabs}
\usepackage{threeparttable}
\usepackage{multirow}
\usepackage{morefloats}
\usepackage[nolists, tablesfirst, nomarkers]{endfloat}
\usepackage{natbib}
\usepackage{amsmath}
\usepackage{setspace} 
\usepackage[footnotesize]{caption}
\usepackage{longtable}
\let\normallongtable=\longtable
\renewcommand{\longtable}{\clearpage\normallongtable}
\DeclareDelayedFloatFlavour*{longtable}{table} 

\usepackage{comment}
\usepackage{appendix}
\usepackage{hyperref} 
\hypersetup{colorlinks=true,
            citecolor=blue,
            linkcolor=blue,
            urlcolor=blue,
            }
\usepackage{rotating}
\usepackage{pdflscape}
\usepackage{pdfpages}
\def\sym#1{\ifmmode^{#1}\else\(^{#1}\)\fi}
\renewcommand{\baselinestretch}{1.5}

\usepackage{natbib}

\begin{document}


\title{Political regime and COVID 19 death rate: efficient, biasing or simply different autocracies?} \author{Guilhem Cassan\thanks{University of Namur, CEPR, DEFIPP, CRED and CEPREMAP. Email: guilhem.cassan@unamur.be} and Milan Van Steenvoort\thanks{Maastricht University. Email: m.vansteenvoort@maastrichtuniversity.nl  \newline
We are grateful to Jeremie Decalf, Romain Lutaud, Glenn Magerman, Marc Sangnier and Vincenzo Verardi for helpful discussions and suggestions. We thank seminar participants at UNamur. Guilhem Cassan thanks CEPREMAP and the FNRS for financial support. Research on this project was financially supported by the Excellence of Science (EOS) Research project of FNRS O020918F. All errors remain our own.}}

\maketitle
\renewcommand{\baselinestretch}{1.5}

\begin{abstract}
The difference in COVID 19 death rates across political regimes has caught a lot of attention. The ``\emph{efficient autocracy}'' view suggests that autocracies may be more efficient at putting in place policies that contain COVID 19 spread. On the other hand, the ``\emph{biasing autocracy}'' view underlines that autocracies may be under reporting their COVID 19 data. We use fixed effect panel regression methods to discriminate between the two sides of the debate. Our results show that a third view may in fact be prevailing: once pre-determined characteristics of countries are accounted for, COVID 19 death rates equalize across political regimes. The difference in death rate across political regime seems therefore to be primarily due to omitted variable bias.
\end{abstract}


\input{introduction}

\input{data}

\input{method}

\input{results}

\input{controls_by_controls}

\input{discussion}


\bibliography{../Biblio/References}
\bibliographystyle{apalike}


\appendix
\input{appendix}

\end{document}

%% file: introduction.tex
\newpage
\section{Introduction}
\begin{figure}[h!]
    \caption{Evolution of COVID 19 data reporting by political regime, time since first 0.4 cases per 100,000}
    \label{fig:DemocraticVsAutocratic}
            \includegraphics[width=\textwidth]{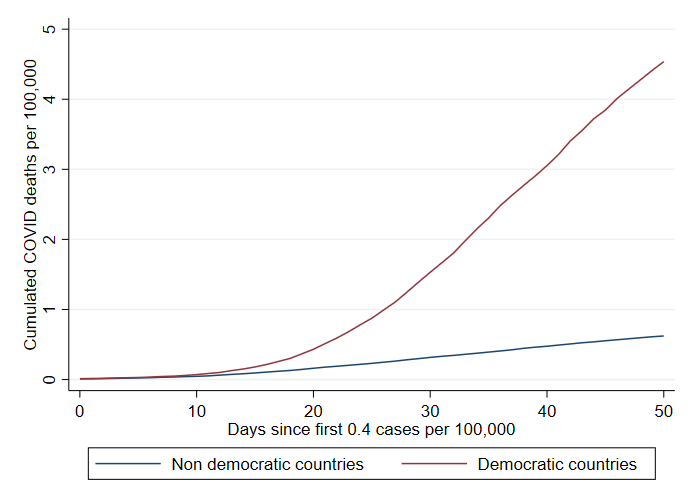}
        \caption*{Total deaths per 100,000. Country level average per type of regime. Democratic countries are defined as a country with a polity score $>$ 0.} 
\end{figure}

While democratic countries have previously been shown to overperfom compared to autocracies with respect to health outcomes \citep{Francoetal2004,BesleyKudamatsu2006,Kudamatsu2012,BOLLYKYetal2019,PIETERSetal2016}, data shows that, in the specific case of the COVID 19 pandemic, democratic countries may be fairing much worse \citep{Sorcietal2020a}, as illustrated in Figure \ref{fig:DemocraticVsAutocratic}.
Indeed, 50 days after the beginning of the pandemic\footnote{Defined as when the number of cases reaches 0.4 per 100,000 in a country.}, democratic countries' COVID 19 death rate is on average larger than that of non democratic countries by approximately 3.9 per 100,000. That is, 50 days after the beginning of the pandemic in a democratic country, the fatality rate in a democracy is on average 7.3 times larger than in an autocracy. \\

A debate \citep{Ang2020} has emerged trying to unpack the reasons behind such wide differences across political regimes: a priori, \emph{all other things equal}, the political regime should not be related to the spread of a disease. We distinguish three main hypothesis to explain this difference.\\

A first interpretation relates to the relative efficiency of social distancing measures in democracies and autocracies. Some have argued that democracies may be less well equipped to implement and enforce social distancing policies \citep{Cepalunietal2020,Sorcietal2020a}, or that they may be implementing them with a sub optimal timing \citep{Sebhatuetal2020}. That is, in this view, autocracies are more able to implement social distancing measures. We will refer to this interpretation as the \emph{efficient autocracy} hypothesis. \\

A second interpretation is that there may be \emph{voluntarily} misreporting of COVID 19 data, in particular by non democratic countries. For example, \cite{Tuiteetal2020a} report that Egypt may have underreported its number of cases, \cite{Tuiteetal2020b} report that Iran may also have underreported its number of cases, while \cite{KAVANAGH2020} discusses that China's political regime may have hindered its initial response to the pandemic. In this view, there are systematic differences between the \emph{real} and the \emph{reported} death rate. These differences are \emph{voluntary} and systematically linked to the type of political regime. We will refer to this interpretation as the \emph{biasing autocracy} hypothesis.\\

A third interpretation has caught less attention \citep{Ashraf2020}: democracies and autocracies tend to have systematically different characteristics apart from their political regimes. These differences, once accounted for, may in fact be sufficient to explain the difference in both the \emph{real} and \emph{reported} death rate. This would leave the contributions due to \emph{voluntary} under-reporting or differences in policies to matter only marginally. An example of such differences would be that autocracies tend to have much younger populations (and therefore, a much smaller \emph{real} death rate, all other things equal) but also a lower ability to test (and therefore, a much smaller \emph{reported} death rate, all other things equal). We refer to this interpretation as the \emph{simply different autocracy} hypothesis. \\

The three aforementionned hypotheses are not mutually exclusive, and simple reduced form econometric approaches can help measuring how much each of them contributes to explain the differences observed across political regimes.
Take the case where the econometrician only observes a \emph{reported} death rate rather than the \emph{real} death rate but can observe the variables determining COVID 19 \emph{real} and \emph{reported} death rate. Also assume that there are two such types of variables: fixed characteristics\footnote{We call fixed characteristics the variable that are pre determined and can not be changed in the time horizon of interest in the paper. In the long run, all characteristics determining the death rate such as, say, the GDP per capita, can of course be considered at least partly as an outcome of the political regime \citep{Acemogluetal2019}.} (say, the share of the population aged 65+ who would determine real death rate or the number of hospital beds per capita who would determine both real and reported death rates) and policy response.
Under the \emph{efficient autocracy} hypothesis, regressing the \emph{reported} death rate on a measure of democracy and controlling for all fixed parameters would lead to a positive and significant coefficient on democracy. However, further controlling for policy response in the regression should bring the coefficient on democracy close to zero and render it non significant. That is, all the differences observed between democracies and autocracies in their \emph{reported} death rate, once fixed characteristics are accounted for, would be due to the difference in policy response across these two types of regime. In this case, there may be a difference between the \emph{real} and the \emph{reported} death rate, but this difference is not systematically linked to the political regime. In fact, these results would indicate that the policy response of autocracies is better than that of democracies, from the perspective of COVID 19 death rate.\\

Under the \emph{biasing autocracy} hypothesis, in a regression of \emph{reported} death rates on a measure of democracy and all relevant controls (including policy response), the coefficient on democracy should be positive and significant. That is, despite controlling for all relevant characteristics and policy response, there still is a systematic difference between democratic and non democratic countries which is not accounted for. In that case, the only reason why a difference may remain would be due to systematic underreporting of casualties by non democratic regimes. This would be due to the fact that the difference between the \emph{real} and the \emph{reported} death rate is always larger for autocracies.\footnote{Note that our methodology is neutral with respect to which political regime may be biasing and allows for democratic regimes to be underreporting more than non democratic regimes.} While the \emph{real} death rate would be identical once all confounding factors are accounted for, the \emph{reported} death rates remain different \emph{even when controlling for the characteristics influencing non voluntary under reporting}. \\

We use daily level data of COVID 19 death rates of 137 countries for the first 50 days of the epidemic and resort to simple reduced form econometric methods using the panel structure of the data. 
First we start by looking into the evolution of daily total death rates across political regimes, using a regression with no controls except country fixed effects (Regression 1). We then include controls for fixed characteristics of countries that are likely to determine the real COVID 19 death rate and allow them to matter differently across time (Regression 2). Finally, we also include controls for the stringency of social distancing measures and allow these to matter differently across time (Regression 3). Comparing Regression 2 to Regression 3 addresses the \emph{efficient autocracy} hypothesis: any difference between the coefficient on democratic regime between Regressions 2 and 3 would be due to the differential in policy response across political regimes. An increase would indicate that autocracies implement more stringent social distancing measures that are successful in decreasing the death rate. Comparing Regression 1 to Regression 3 addresses the \emph{biasing autocracy} and the \emph{simply different autocracy} hypothesis: once all controls for both fixed characteristics and policy response are accounted for, does the difference between autocratic and democratic regimes remains (\emph{biasing autocracy} hypothesis) or vanishes (\emph{simply different autocracy} hypothesis)? \\

Our results indicate that the inclusion of controls for country characteristics and policy response is in fact enough to remove almost all cross regime difference in COVID 19 mortality rates. In particular, the population susceptibility to die upon contamination and geographical characteristics seem to be the main contributors explaining the differences observed across political regimes. \\

Section 2 of this paper presents the data. Section 3 elaborates on the methodology used to test our hypotheses. Section 4 and 5 present our main results. Finally, Section 6 provides a discussion of our main findings and Section 7 concludes. \\

%% file: data.tex
\section{Data}
In order to investigate our hypothesis, we assemble a dataset that comprises information on daily cases and deaths in the first 50 days of the pandemic for 137 countries. Our dependent variable, the daily country-level total number of reported cases and reported deaths due to the COVID-19 virus is from \cite{DONGetal2020} \footnote{Last accessed: 16.10.2020}. Our main variable of interest, the classification of political regimes along the autocratic-democratic scale, comes from the Polity IV project \citep{PolityIV}. \\

Under the \emph{simply different autocracy} hypothesis, accounting for the differences in characteristics of countries would suffice to explain the difference in reported mortality rates across political regimes. We therefore collected an extensive array of country level variables. To proxy for income and health infrastructure differences, we gathered data on gross domestic product per capita in 2018 from the World economic outlook survey (IMF), and completed it with the World Factbook (CIA). Furthermore, information on the number of available hospital beds (per thousand inhabitants) is retrieved from the World Bank, to account for differences in health infrastructure that may drive the mortality difference (actual and reported death rates). \\

To capture differences in demographic characteristics which may explain the speed of the spread of the disease, we use data on countries' total population and density in 2019 from the World Bank, and data on countries' urbanization rate in 2019 from the World in Data website. To control for the effect of geographical characteristics, we collect data on the latitude and longitude of each country's capital from the World Cities Database, and classify each country according to its World Bank region.\footnote{These are: East Asia and Pacific, Europe and Central Asia, Latin America and Caribbean, Middle East and North Africa, North America, South Asia and Sub-Saharan Africa.}
Finally, to control for population risk of mortality, we include the share of population aged 65+ (from the World Bank) and, since air pollution has been shown to be associated with COVID 19 death rates \citep{Zhuetal2020}, we use summary exposure values to ambient ozone pollution and ambient particle matter pollution from the Global Burden Disease dataset (2017). \\

 To test the \emph{efficient autocracy} hypothesis, we use information on countries' different COVID 19 containment policies from the ``Variation in Government Responses to COVID-19'' dataset \citep{Haleetal2020}. This dataset includes a daily policy stringency index based on the aggregation of 17 policy indicators\footnote{Policy indicators include policies with respect to closures or movement restrictions as well as economic and health system policies. Last accessed: 16.10.2020.}.\\

Given that the data on our dependent variable is at the daily level, this allows us to construct a panel dataset that comprises a total of 137 countries\footnote{See Appendix \ref{app:countrypolity} for the list of the countries present in our dataset and their classification as democratic or non democratic.}, classified as either democratic or non democratic, for which we have information on all the previously mentioned national characteristics. Therefore, our dataset displays information (by day and by country) on the total number of reported deaths due to the COVID-19 virus, on the stringency of policy measures taken by a given country, and on all other relevant characteristics of that country. We focus on the first 50 days since the beginning of the pandemic in each country, which we define as having more than 0.4 cases per 100,000 \footnote{Because our outcome of interest is death per capita, it makes sense to use cases per capita rather than the absolute number of cases to determine the beginning of the pandemic. In Appendix \ref{app:rob_threshold}, we show that results are robust to using alternatives thresholds. In Appendix \ref{app:pandemic_start}, we show that the timing of the beginning of the pandemic does not seem to differ significantly across political regimes.}\\

%% file: method.tex
\section{Methodology}
Given the country-day panel structure of our data, we resort to fixed effect panel reduced form econometric methods to look into the differences in COVID 19 casualty rates across political regimes and time. This method allows us to remove the influence of all time invariant differences across countries by including countries fixed effects. This further allows us to control for an extensive set of countries' pre-determined characteristics and for differences in containment policies across countries.\\

We specify the following regression equation, which we run using Ordinary Least Squares: 

   \begin{equation}\label{eq:2}
   \begin{array}{lll}
        DeathRate_{ct} & = & \sum_{t=1}^{T} \beta_t * \mathrm{democratic}_{c} * time.from.start_t \\
        			   & + & \sum_{t=1}^{T} \alpha_t * time.from.start_t  \\
        			   & + & \sum_{t=1}^{T} \delta_t * \mathrm{X}_{c} * time.from.start_t  \\
        			   & + & \sum_{t=1}^{T} \gamma_t * \mathrm{Y}_{ct} * time.from.start_t + \mathrm{\delta}_{c} + \omega_{ct} \\
        			   & + & \epsilon_{ct}  
    \end{array}
    \end{equation}

$\mathrm{DeathRate}_{ct}$ is the log of daily declared total deaths per 100,000 inhabitants in country c plus one, t days after the beginning of the pandemic in country c. $\mathrm{democratic}_{c}$ is a dummy indicating that the Polity IV score of country c is positive. $time.from.start_t$ is a set of fixed effect for each day since the beginning of the pandemic. The interaction of $\mathrm{democratic}_{c}$ with $time.from.start_t$ allows us to track day by day the evolution of the difference in death rates across political regimes, a standard approach in economics (see \cite{duflo01} or \cite{Cassan2019} among others).\\

$X_{c}$ is a large set of controls for countries' pre determined characteristics: GDP per capita, number of hospital beds per 1000, population, density, urbanization rate, share of population aged 65+, summary exposure value to particle matters pollution, summary exposure value to ambient ozone pollution, as well as for the square of these variables, World Bank regions fixed effects, latitude and longitude. We interact all these variables with the $time.from.start_t$ fixed effects to allow their effect to vary over time. \\

$Y_{ct}$ is a measure of country policy response to the pandemic. It is a stringency index of governmental response (as measured at t-15 to allow for lags in its effect) We also include the square level of this variable to allow for non-linear effects. Furthermore, we interact these variables with the set of $time.from.start_t$ fixed effects, to control for their time varying effect.
$\mathrm{\delta}_{c}$ is a set of country fixed effects. Finally, $\omega_{ct}$ is a set of day of the week fixed effect interacted with $time.from.start_t$ fixed effects, to control for variations in reporting across days of the week. \\

We perform this regression iteratively. First, we do not implement any of the $X_{c}$ and $Y_{ct}$ controls (Regression 1). This allows us to see the evolution of the difference in casualty rates across political regimes when no confounding factors are accounted for. Then, we implement $X_{c}$ but not $Y_{ct}$ (Regression 2). This will allow us to see how much of the difference across political regimes survives once the different pre-determined characteristics of countries are accounted for. Finally, we add the $Y_{ct}$ policy response controls (Regression 3).\\

This iterative procedure allows us to address the different sides of the debate on the role of political regime in fighting COVID 19. Comparing Regression 2 to Regression 3 addresses the \emph{efficient autocracy} hypothesis: any difference between the $\beta_t$ coefficients on democratic regime between Regressions 2 and 3 would be due to the differential in policy response across political regimes. An increase would indicate that autocracies implement more stringent social distancing measures that are successful in decreasing the COVID 19 death rate. \\

Comparing Regression 1 to Regression 3 addresses the \emph{biasing autocracy} and \emph{simply different autocracy} hypothesis: once all controls for both fixed characteristics and policy response are accounted for, do the $\beta_t$ coefficients remain positive (\emph{biasing autocracy}) or do they equalize to zero (\emph{simply different autocracy})? Note that these hypotheses are not mutually exclusive: autocracies may well be efficient, biasing and different at the same time. Our methodology allows to capture this possibility: if the $\beta_t$ coefficients decrease but remain large and significant when passing from Regression 1 to Regression 3 and change but remain large and significant between Regression 2 and Regression 3, then this would support the simultaneous presence of the three hypotheses.

%% file: results.tex
\section{Results}

\begin{figure}[h!]
    \caption{Evolution of COVID 19 log death per 100,000 since 0.4 cases per 100,000. 95\% CI.}
    \label{fig:DataSinceOnset4}
        \begin{minipage}[b]{0.45\linewidth}
            \centering
            \includegraphics[width=\textwidth]{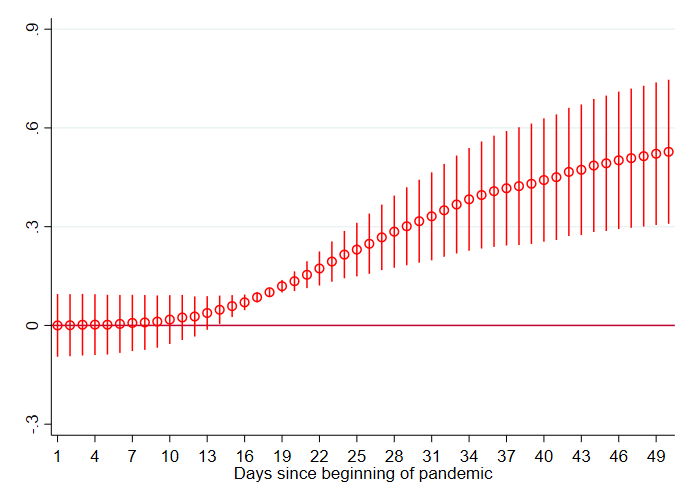}
            \caption*{No Controls}
        \end{minipage}
        \hspace{0.5cm}
        \begin{minipage}[b]{0.45\linewidth}
            \centering
            \includegraphics[width=\textwidth]{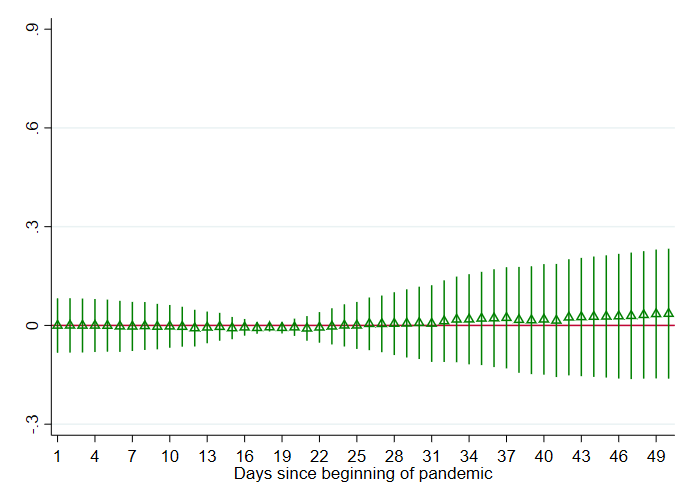}
            \caption*{Controlling for pre-determined characteristics}
        \end{minipage} 
        \hspace{0.5cm} 
        \begin{minipage}[b]{0.45\linewidth}
            \centering
            \includegraphics[width=\textwidth]{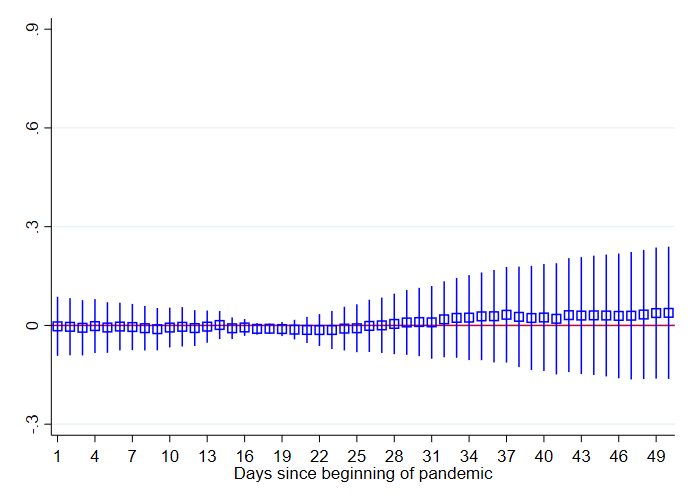}
            \caption*{Controlling for pre-determined characteristics and policy response}
        \end{minipage} 
    \caption*{Standard errors are two way clustered at the country and day level.}
\end{figure}

Figure \ref{fig:DataSinceOnset4} presents the $\beta_t$ coefficients from Equation \ref{eq:2} for all three versions of the specification.\footnote{In Appendix \ref{app:rob_threshold}, we show that results are robust to using alternatives definition of the start of the pandemic in a country.}
The first panel presents the results of Regression 1, when no controls excepting country and day of the week fixed effects are included. The death rates start diverging across political regimes roughly 10 days after the beginning of the pandemic. After 50 days, the $\beta_{t}$ coefficient reaches 0.5, which represents 135\% of the mean. \\

The second panel includes controls for pre-determined characteristics interacted with day fixed effects. The $\beta_{t}$ coefficients become precisely estimated zeros. That is, once countries' differences in characteristics are taken into account, the difference in death rates across political regimes does not survive. Hence, we do not find support for the \emph{biasing autocracy} hypothesis with our methodology. \\

The third panel adds controls for countries' policy response to the pandemic. As a result, our coefficients of interest $\beta_{t}$ remain virtually unaffected. That is, our results do not support the \emph{efficient autocracy} hypothesis.\footnote{To test if these results are driven by an outlier country, we run Regression 1, 2 and 3 137 times, removing one country at a time. We plot the 50 coefficients of interests of each of these 411 regressions in Figure \ref{fig:RemovingCountries} of Appendix \ref{app:removing_country}. It can be seen that the results are robust to the omission of any single country.} 
Therefore, once systematic differences across countries' characteristics and policy responses are taken into consideration, the differences in death rates apparent in Figure \ref{fig:DemocraticVsAutocratic} and in the first panel of Figure \ref{fig:DataSinceOnset4} vanish. The reason why reported COVID 19 death rates differ across political regimes is fully accounted for by factors which systematically differ between democracies and autocracies. That is, our results indicate that the \emph{simply different autocracy} hypothesis is prevailing. That is, our results do not support neither the fact that autocracies are more efficient at controlling the pandemic nor that they are voluntarily under reporting casualty more often.\\

%% file: controls_by_controls.tex
\section{Which characteristics matter?}
\begin{figure}[h!]
    \caption{Coefficient on ``Democracy'' and introduction of controls}
    \label{fig:CoefficientDemocracy}
        \begin{minipage}[b]{0.45\linewidth}
        \centering
        \includegraphics[width=\textwidth]{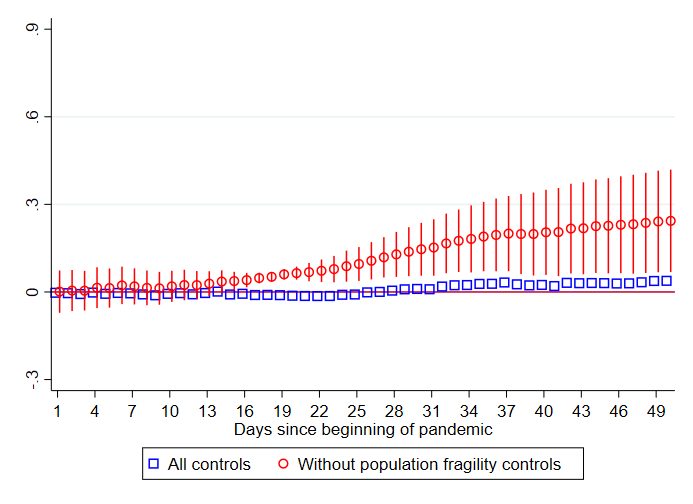}
        \caption*{No Population fragility controls}
        \end{minipage} 
        \hspace{0.5cm} 
        \begin{minipage}[b]{0.45\linewidth}
            \centering
            \includegraphics[width=\textwidth]{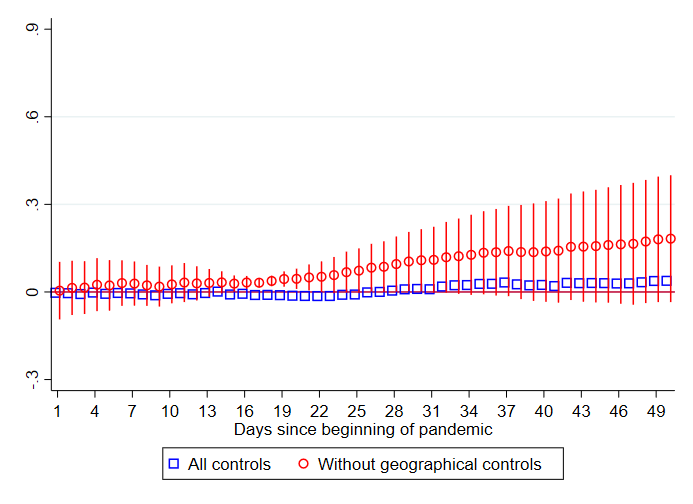}
        \caption*{No Geographic controls}
        \end{minipage} 
        \hspace{0.5cm}
        \begin{minipage}[b]{0.45\linewidth}
            \centering
            \includegraphics[width=\textwidth]{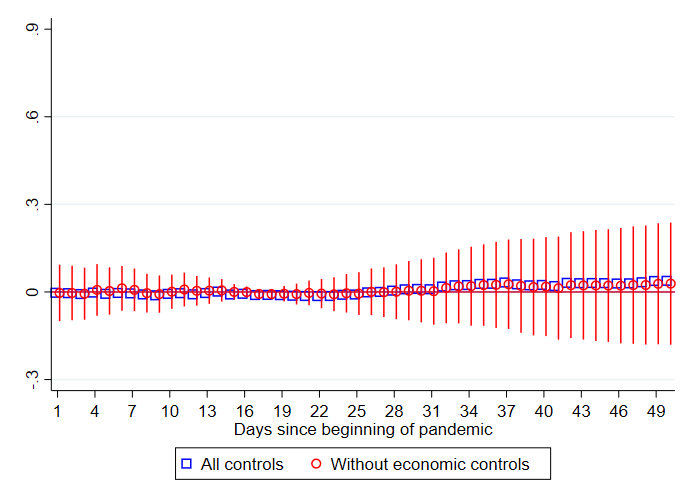}
        \caption*{No Economic controls}
        \end{minipage} 
        \hspace{0.5cm}
        \begin{minipage}[b]{0.45\linewidth}
            \centering
            \includegraphics[width=\textwidth]{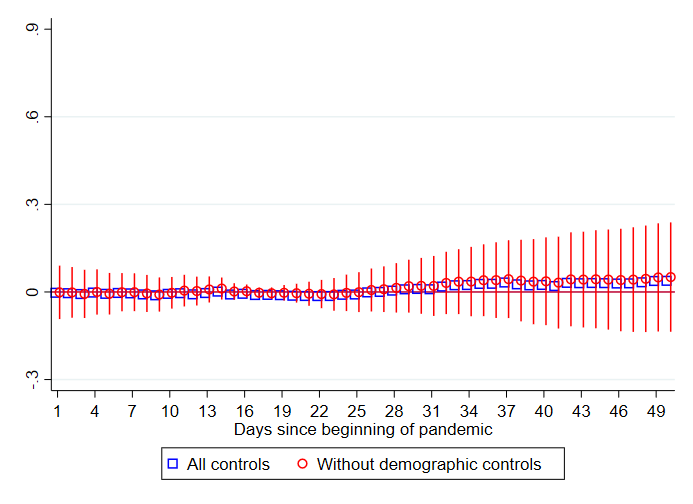}
        \caption*{No Demographic controls}
        \end{minipage} 
        \hspace{0.5cm} 
        \begin{minipage}[b]{0.45\linewidth}
            \centering
            \includegraphics[width=\textwidth]{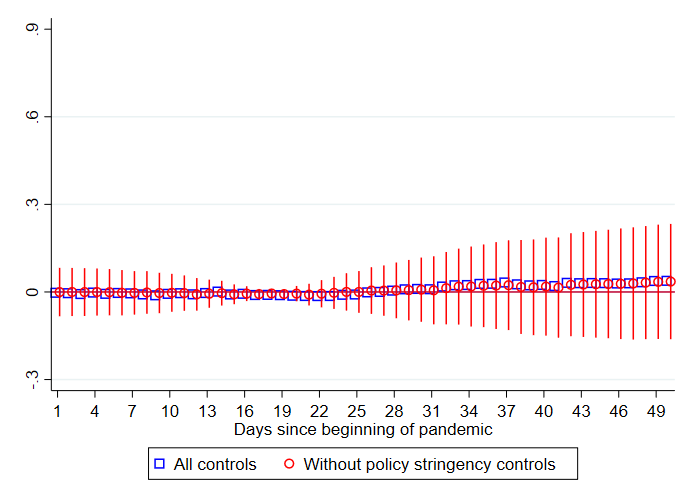}
        \caption*{No Policy Response controls}
        \end{minipage} 
    \caption*{Standard errors are two way clustered at the country and day level.}
\end{figure}

Having seen that the inclusion of controls is sufficient to remove the ``political regime'' effect on COVID 19 death rate, we now move to a related question: which characteristics are contributing to closing the COVID 19 death rate gap between autocratic and democratic gap?
In order to do so, we group our control variables in five categories:\\
- Geographical controls (latitude, longitude, World Bank region fixed effects)\\
- Wealth controls (GDP per capita, hospital beds per capita). These will proxy for the quality of the health system in the country, and will likely influence both the \emph{real} and the \emph{reported} death rate. \\
- Demographic controls (population, density, urbanization rate): these are likely to influence the speed of the spread of the pandemic. \\
- Population fragility controls (share of population aged 65+ and exposure to pollution): these are likely to influence the lethality of COVID 19 for a given spread of the disease. \\
- Policy response. \\
We run Regression \ref{eq:2}, removing each of these groups of controls one at a time. Figure \ref{fig:CoefficientDemocracy} presents the $\beta_t$ coefficients for each of these regressions. It also includes the coefficients of the original results of Regression \ref{eq:2} for comparison (we do not report the confidence intervals on these coefficients for readability). \\

Two set of controls seem to matter most. The first set of controls are the controls related to population fragility (which contain notably the share of population aged 65+). The second set of controls that seems to affect the coefficient is the geographic controls set.
That is, the main reason why mortality rates from COVID seem to differ across political regimes may be that autocratic regime tend to have a population that is less susceptible to die from COVID 19 and to be located in regions in which COVID 19 appears to be less lethal. Interpreting the geographic controls is not straightforward, but it is reasonable to say that they seem to indicate that country characteristics correlated with geography and not captured by our extensive set of controls may play an important part in explaining COVID 19 mortality.

%% file: discussion.tex
\section{Discussion}
A few remarks are in order to help interpret our results. First, one should keep in mind that the variables that we consider pre-determined characteristics, such as the GDP per capita are only pre-determined in the time horizon that we are considering. Over the long run, they are an outcome of the political regime. See for example \cite{Acemogluetal2019}, who show that democracy causes growth. In that sense, our results do not take into consideration the long term effect of political regimes on the variables that may determine COVID 19 death rates. \\

For instance, a better health care system will lead to both a lower \emph{real} death rate (infected individuals are better treated) and a higher \emph{reported} death rate (infected individuals' death is better attributed to COVID 19). If, as has been argued in the literature \citep{Francoetal2004,BesleyKudamatsu2006,Kudamatsu2012,BOLLYKYetal2019,PIETERSetal2016}, democracies tend to have better health care policies; in the long run, the health care system (which we consider as pre-determined) will be better in democracies because of the political regime, which will causally affect both \emph{real} and \emph{reported} death rates across political regimes. \\

Second, our focus is only on COVID 19 death rates. Arguably, however, one may have wanted to study death rates from all causes rather than just from COVID 19. Even in times of pandemic, governments should aim at preserving the health of their citizens from all sources of harm, not from one specific cause. Given the attention given to COVID 19 death rates, a pro-democracy argument would be that while there does not seem to be differences across political regimes for COVID 19 death rates, this may hide the fact that autocracies have focused on decreasing COVID 19 death rate at the expense of death from other sources. \\

One could develop this idea even further and argue that democracies have higher COVID 19 mortality rates because they are better at preventing non COVID 19 deaths, leading to a population which is on average older and therefore more likely to die if infected by COVID 19. This question can unfortunately not be tackled with the available data, and we leave it to future research (when mortality data from all causes will be available for a sufficient number of countries), but note that the differences in countries' population's susceptibility to die from COVID 19 upon contamination seem to be one of the main drivers of the difference in COVID 19 mortality rates across political regimes, all other things equal.\\

Third, our findings do not contradict previous studies on under reporting of COVID 19 data, in particular country specific studies. Indeed, because of the statistical analysis used, our results do not imply that no single country underreported or manipulated its COVID 19 mortality data. However, our results do address the widespread idea that autocracies are systematically and willingly under reporting COVID 19 casualties. What our results do indicate is that under reporting (by any political regime) is primarily due to the different characteristics of countries that are correlated with the political regime rather than a direct causal effect of the political regime. \\

That is, a plausible interpretation is that autocratic governments may well under report data while not manipulating it. One could argue that even if autocracies are under reporting COVID 19 death rates, this may be primarily driven by their overall incapacity to link death to its cause rather than to a direct attempt at data manipulation. The low reported COVID 19 death rate in autocracies may in part be due to the lower level of development of both the public health infrastructure and the statistical apparatus of autocracies. However, and this goes back to our first point, over the long run, public health infrastructure and statistical apparatus may well be determined by the political regime. \\

\section{Conclusion}
In this paper, we investigated the COVID 19 death rate gap between democratic and autocratic countries. We formulated three main hypotheses based on the previous literature: the gap can be due to the fact that autocracies are more efficient at implementing restricting policy measures; that autocracies are underreporting their COVID 19 data and that autocracies simply have different characteristics that can explain the death rate gap.
Our analysis, relying on simple econometric tools, allows to make progress in the debate around the sources of the observed differences in COVID 19 death rates across political regimes. \\

We show that once pre-determined characteristics and policy responses are taken into account, COVID 19 death rates do not exhibit any difference across political regimes: the coefficients on democracy become precisely estimated zeros. Our results therefore do not show support neither for the \emph{efficient autocracy} nor for the \emph{biasing autocracy} hypotheses, as we do not find evidence that autocracies are neither systematically better at preventing COVID 19 death nor that they are more often under reporting casualties. \\

Our findings indicate that democracies and autocracies are \emph{simply different}, and that these differences are sufficient to account for all the observed differences in COVID 19 death rates. Finally, we analyzed which controls matter most in explaining the difference in COVID 19 death rates. We found that characteristics related to the vulnerability of the population to the disease and geographical controls appeared to be of significant importance. \\

%% file: appendix.tex
\section{Appendix}

\subsection{Data sources and descriptive statistics}
\label{app:data_sources}

Table \ref{tab:datasources} presents the data sources used to compute the variables exploited in our analysis.

\begin{sidewaystable}[h!]
    \begin{center}
        \caption{Datasources.}
        \label{tab:datasources}
        \begin{tabular}{l|l}
            \toprule
            \textbf{Data} & \textbf{Source}\\
            \midrule
            COVID19 Death Rate & Dong et al. (2020) \\
            COVID19 Cases & Dong et al. (2020) \\
            Democratic & Polity IV project (Center for Systemic Peace, 2015) \\
            Stringency Index of Policy Response & Variation in Government Responses to COVID-19 (Hale et al., 2020). \\
            Gross Domestic Product per capita & World Economic Outlook, IMF (2018) + World Factbook, CIA (2018) \\
            Share of 65+ & World Bank (2019) \\
            Population Density & World Bank (2019)  \\
            Population & World Bank (2019) \\
            Urbanization Rate & World in Data \\
            Hospital Beds per 1000 & World Bank \\
            Summary Exposure Value  to Air Pollution & Global Burden of Disease (2017) \\
            Summary Exposure Value  to Ambiant Ozone Pollution & Global Burden of Disease (2017) \\
            Latitude and Longitude & World Cities Database \\ 
            World Regions & World Bank \\
            \bottomrule
        \end{tabular}
    \end{center}
\end{sidewaystable}

\begin{sidewaystable}
    \centering 
    \caption{Descriptive statistics 
    \label{tab:des_stat}}
    \input{Tables/stat_des}
\end{sidewaystable}

Table \ref{tab:des_stat} presents the descriptive statistics for our variable. Our final dataset comprises a total of 137 countries for which we have all the variables and for which we observe 50 days of data since the beginning of the pandemic.

\subsection{Polity score, by country}
\label{app:countrypolity}
Table \ref{tab:polity_score} presents the polity IV score of all countries that are included in our sample. Countries whose score is higher than 0 are classified as democratic.

\begin{longtable}{cccc}
    \caption{Polity Score, by country} 
    \label{tab:polity_score} \\
    \toprule
   \multicolumn{1}{c}{Country} &\multicolumn{1}{c}{Polity Score} &\multicolumn{1}{c}{Country} &\multicolumn{1}{c}{Polity Score} \\ \midrule
    \endfirsthead

    \toprule
   \multicolumn{1}{c}{Country} &\multicolumn{1}{c}{Polity Score} &\multicolumn{1}{c}{Country} &\multicolumn{1}{c}{Polity Score}  \\ \midrule
    \endhead

    \bottomrule
    \endfoot

    \bottomrule
    \endlastfoot

Afghanistan &   -1  &   Kuwait  &   -7  \\
Albania &   9   &   Kyrgyz Republic &   8   \\
Algeria &   2   &   Latvia  &   8   \\
Argentina   &   9   &   Lebanon &   6   \\
Australia   &   10  &   Liberia &   7   \\
Austria &   10  &   Libya   &   -7  \\
Azerbaijan  &   -7  &   Lithuania   &   10  \\
Bahrain &   -10 &   Luxembourg  &   10  \\
Bangladesh  &   -6  &   Madagascar  &   6   \\
Belarus &   -7  &   Malawi  &   6   \\
Belgium &   8   &   Malaysia    &   7   \\
Benin   &   7   &   Mali    &   5   \\
Bhutan  &   7   &   Mauritius   &   10  \\
Bolivia &   7   &   Mexico  &   8   \\
Botswana    &   8   &   Moldova &   9   \\
Brazil  &   8   &   Mongolia    &   10  \\
Bulgaria    &   9   &   Morocco &   -4  \\
Burkina Faso    &   6   &   Mozambique  &   5   \\
Burundi &   -1  &   Myanmar &   8   \\
Cabo Verde  &   10  &   Nepal   &   7   \\
Cambodia    &   -4  &   Netherlands &   10  \\
Cameroon    &   -4  &   New Zealand &   10  \\
Canada  &   10  &   Nicaragua   &   6   \\
Central African Republic    &   6   &   Niger   &   5   \\
Chile   &   10  &   Nigeria &   7   \\
China   &   -7  &   Norway  &   10  \\
Colombia    &   7   &   Oman    &   -8  \\
Costa Rica  &   10  &   Pakistan    &   7   \\
Croatia &   9   &   Panama  &   9   \\
Cuba    &   -5  &   Paraguay    &   9   \\
Cyprus  &   10  &   Peru    &   9   \\
Czech Republic  &   9   &   Philippines &   8   \\
Denmark &   10  &   Poland  &   10  \\
Djibouti    &   3   &   Portugal    &   10  \\
Dominican Republic  &   7   &   Qatar   &   -10 \\
Ecuador &   5   &   Romania &   9   \\
Egypt, Arab Rep.    &   -4  &   Russian Federation  &   4   \\
El Salvador &   8   &   Saudi Arabia    &   -10 \\
Estonia &   9   &   Singapore   &   -2  \\
Eswatini    &   -9  &   Slovakia    &   10  \\
Ethiopia    &   1   &   Slovenia    &   10  \\
Fiji    &   2   &   Spain   &   10  \\
Finland &   10  &   Sri Lanka   &   6   \\
France  &   9   &   Sudan   &   -4  \\
Gabon   &   3   &   Suriname    &   5   \\
Gambia, The &   4   &   Sweden  &   10  \\
Georgia &   7   &   Switzerland &   10  \\
Germany &   10  &   Syrian Arab Republic    &   -9  \\
Ghana   &   8   &   Tajikistan  &   -3  \\
Greece  &   10  &   Tanzania    &   3   \\
Guatemala   &   8   &   Thailand    &   -3  \\
Guinea  &   4   &   Timor-Leste &   8   \\
Guyana  &   7   &   Togo    &   -2  \\
Haiti   &   5   &   Trinidad and Tobago &   10  \\
Honduras    &   7   &   Tunisia &   7   \\
Hungary &   10  &   Turkey  &   -4  \\
India   &   9   &   Uganda  &   -1  \\
Indonesia   &   9   &   Ukraine &   4   \\
Iran    &   -7  &   United Arab Emirates    &   -8  \\
Iraq    &   6   &   United Kingdom  &   8   \\
Ireland &   10  &   United States   &   8   \\
Israel  &   6   &   Uruguay &   10  \\
Italy   &   10  &   Uzbekistan  &   -9  \\
Jamaica &   9   &   Venezuela   &   -3  \\
Japan   &   10  &   Vietnam &   -7  \\
Jordan  &   -3  &   Yemen, Rep. &   3   \\
Kazakhstan  &   -6  &   Zambia  &   6   \\
Kenya   &   9   &   Zimbabwe    &   4   \\
Korea, Rep. &   8

\end{longtable}

\subsection{Political regime and start of the pandemic}
\label{app:pandemic_start}
Our analysis focuses on the evolution of the death rates across time since the beginning of the pandemic in each country. We verify whether the political regime determines when these first contaminations are reached, which may imply that the timing that we rely on is biased. In order to do so, we run the following OLS regressions:
 
\begin{align} \label{eq:1} 
    \mathrm{time.to.start}_{c} = \alpha + \beta_{1} * \mathrm{democratic}_{c} + \ \mathrm{X}_{c} + \epsilon_{c} 
\end{align}

 $\mathrm{time.to.start}_{c}$ is the number of days between the start of the pandemic in country c and the $20^{th}$ of January. We specify three variations of Equation \ref{eq:1}: if the country has declared 4 or 6 per 100,000 cases or has reached 100 cases. Table \ref{tab:time_to_start} presents the results. We find no statistically significant difference across political regimes for the start of the pandemic. Note that the number of countries reaching both 100 total cases and for which we observe 50 days since the first 100 case is 135: two countries do not reach 50 days after the first 100 cases in our data compared to our main sample.

 \begin{table}\centering
 \caption{Time to first cases and political regime}
 \label{tab:time_to_start}
 \begin{minipage}[c]{0.80\linewidth}
    {
\def\sym#1{\ifmmode^{#1}\else\(^{#1}\)\fi}
\begin{tabular}{l*{3}{c}}
\toprule
                    &\multicolumn{1}{c}{0.4 c. per 100,000}&\multicolumn{1}{c}{0.6 c. per 100,000}&\multicolumn{1}{c}{100 c.}\\
\midrule
Democracy           &       -1.19         &       -0.68         &        4.11         \\
                    &      (5.91)         &      (6.46)         &      (5.04)         \\
\midrule
R-sq                &        0.58         &        0.57         &        0.57         \\
Observations        &         137         &         137         &         135         \\
\bottomrule
\end{tabular}  
}
{\scriptsize
{\setlength{\baselineskip}{0.9\baselineskip}
Heteroskedasticity-robust standard errors in parentheses * p$<$.10 ** p$<$.05 *** p$<$.01. Controls included are: GDP per capita, population, density, urbanization rate, share of 65 and above, number of hospital beds per capita and the square of all preceding variables, latitude, longitude, World Bank region fixed effect.
\par}
}
\end{minipage}
\end{table}

\subsection{Alternative definitions of the start of the pandemic}
\label{app:rob_threshold}
\begin{figure}[h!]
    \caption{Evolution of COVID 19 log death per 100,000 since 0.6 cases per 100,000. 95\% CI.}
    \label{fig:DataSinceOnset6}
        \begin{minipage}[b]{0.45\linewidth}
            \centering
            \includegraphics[width=\textwidth]{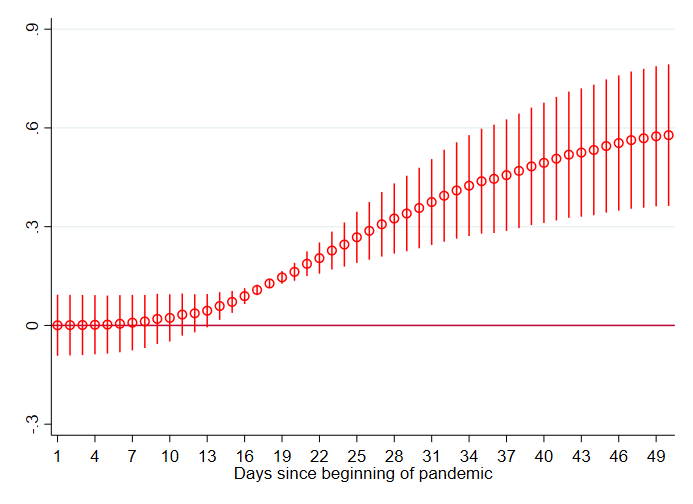}
            \caption*{No Controls}
        \end{minipage}
        \hspace{0.5cm}
        \begin{minipage}[b]{0.45\linewidth}
            \centering
            \includegraphics[width=\textwidth]{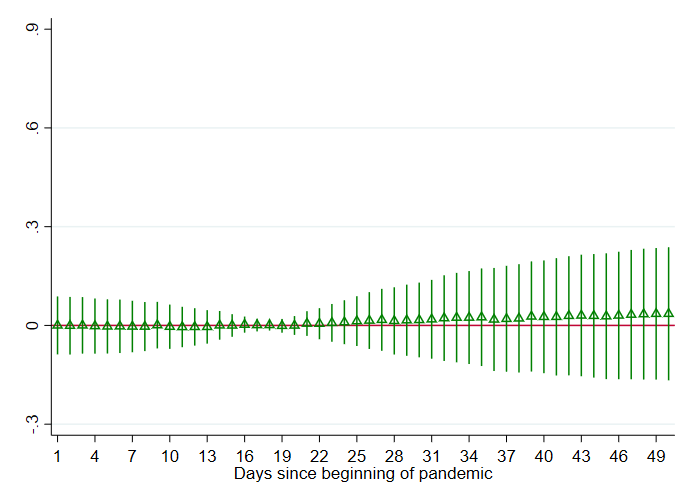}
            \caption*{Controlling for pre-determined characteristics}
        \end{minipage} 
        \hspace{0.5cm} 
        \begin{minipage}[b]{0.45\linewidth}
            \centering
            \includegraphics[width=\textwidth]{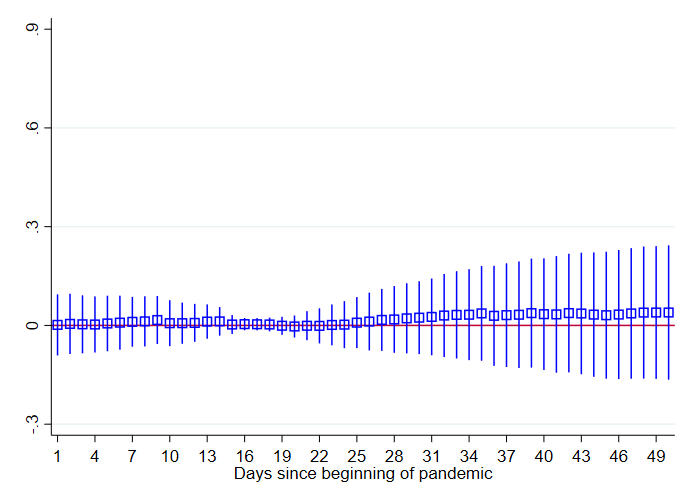}
            \caption*{Controlling for pre-determined characteristics and policy response}
        \end{minipage} 
        \caption*{Standard errors are two way clustered at the country and day level.}
\end{figure}

\begin{figure}[h!]
    \caption{Evolution of COVID 19 log death per 100,000 since 100 cases. 95\% CI.}
    \label{fig:DataSinceOnset100}
        \begin{minipage}[b]{0.45\linewidth}
            \centering
            \includegraphics[width=\textwidth]{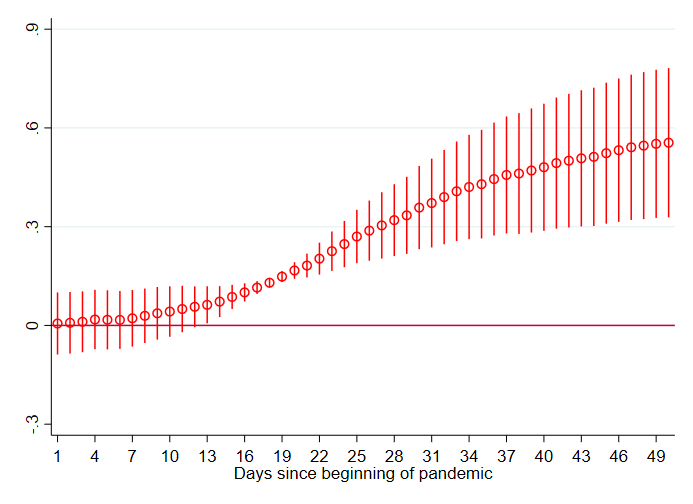}
            \caption*{No Controls}
        \end{minipage}
        \hspace{0.5cm}
        \begin{minipage}[b]{0.45\linewidth}
            \centering
            \includegraphics[width=\textwidth]{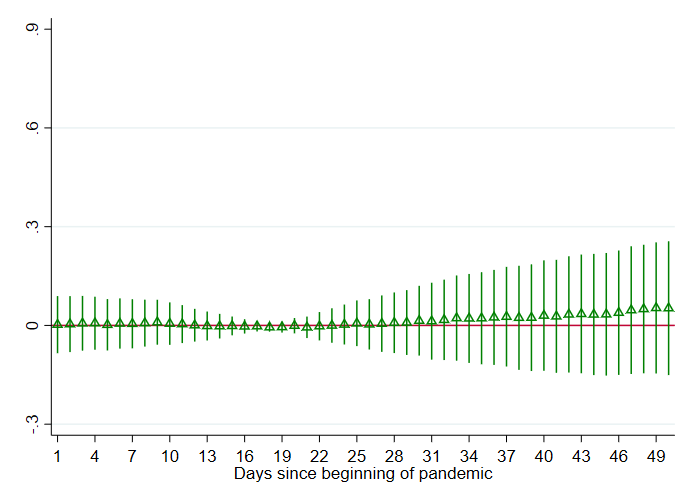}
            \caption*{Controlling for pre-determined characteristics}
        \end{minipage} 
        \hspace{0.5cm} 
        \begin{minipage}[b]{0.45\linewidth}
            \centering
            \includegraphics[width=\textwidth]{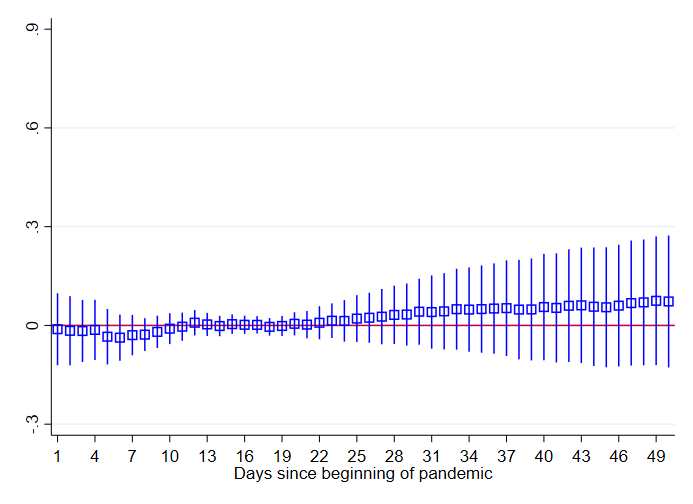}
            \caption*{Controlling for pre-determined characteristics and policy response}
        \end{minipage} 
        \caption*{Standard errors are two way clustered at the country and day level.}
\end{figure}

Figures \ref{fig:DataSinceOnset6} and \ref{fig:DataSinceOnset100} reproduce Figure \ref{fig:DataSinceOnset4}, using 0.6 cases per 100,000 and 100 cases reported cases as the definition of the start of the pandemic in a country, respectively. Results remain unaffected by this change in definition of the beginning of the pandemic. Note that the number of countries reaching both 100 total cases and for which we observe 50 days since the first 100 case is 135: two countries do not reach 50 days after the first 100 cases in our data compared to our main sample.

\subsection{Robustness check: removing one country at a time}
\label{app:removing_country}
\begin{figure}[h!]
    \caption{Removing countries one by one: Evolution of COVID 19 log death per 100,000 since 0.4 cases per 100,000.}
    \label{fig:RemovingCountries}
        \begin{minipage}[b]{0.45\linewidth}
            \centering
            \includegraphics[width=\textwidth]{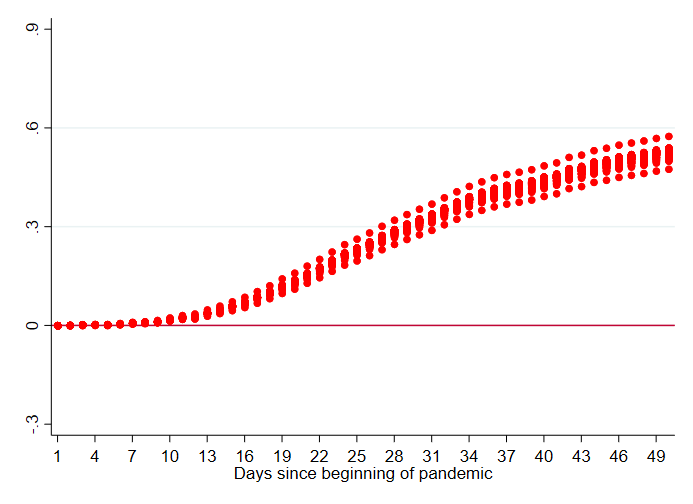}
            \caption*{No Controls}
        \end{minipage}
        \hspace{0.5cm}
        \begin{minipage}[b]{0.45\linewidth}
            \centering
            \includegraphics[width=\textwidth]{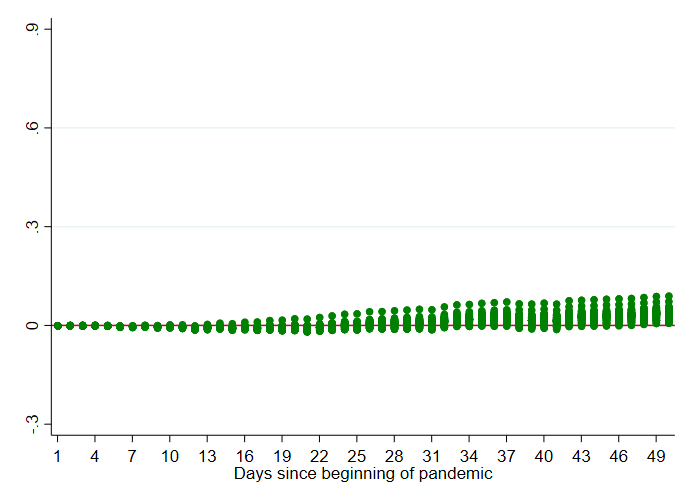}
            \caption*{Controlling for pre-determined characteristics}
        \end{minipage} 
        \hspace{0.5cm}
        \begin{minipage}[b]{0.45\linewidth}
            \centering
            \includegraphics[width=\textwidth]{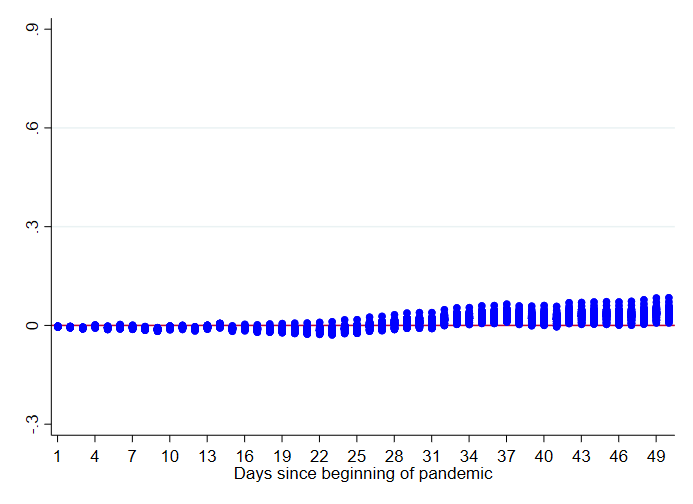}
            \caption*{Controlling for pre-determined characteristics and policy response}
        \end{minipage} 
    \caption*{CI not reported for readability}
\end{figure}

In order to test if an outlier country is driving our findings, we run each regression 133 times, removing one country at a time. 
 We plot the 50 coefficients of interests of each of these 399 regressions in Figure \ref{fig:RemovingCountries}, which replicates Figure \ref{fig:DataSinceOnset4}. It can be seen that the results are robust to the omission of any single country.

%% file: Tables/stat_des.tex
{
\def\sym#1{\ifmmode^{#1}\else\(^{#1}\)\fi}
\begin{tabular}{l*{1}{cccc}}
\toprule
                    &\multicolumn{4}{c}{}                               \\
                    &        Mean&          sd&         Min&         Max\\
\midrule
Democratic          &        0.76&        0.43&        0.00&        1.00\\
Log total deaths per 100,000&        0.37&        0.65&        0.00&        4.08\\
Share of 65+        &        9.56&        6.64&        1.16&       28.00\\
GDP Per Capita      &      23,636&      23,544&         727&     132,886\\
Population in million&       52.97&      170.10&        0.55&    1,397.71\\
Hospital beds per 1000&         3.0&         2.5&         0.1&        13.4\\
Population Density  &       206.9&       779.5&         2.1&     8,829.1\\
Urbanization rate   &          62&          22&          13&         100\\
Summary exposure value to ambient ozone pollution - Age standardized&          36&          13&           2&          53\\
Summary exposure value to ambient particulate matter pollution - Age standardize&          32&          17&           7&          90\\
Stringency t-15     &          63&          29&           0&         100\\
\midrule
Observations        &       6,987&            &            &            \\
\bottomrule
\end{tabular}
}